\documentclass[twocolumn]{aastex631}
\usepackage{graphicx}
\usepackage{amsmath}
\usepackage{amssymb}
\usepackage{color}
\usepackage{hyperref}
\usepackage{listings}

\gdef\xx[#1]{\textcolor{red}{#1}}
\def\code#1{\texttt{#1}}

\newcommand{\GG}[1]{}

\submitjournal{PASP}
\begin{document}

\newcommand\XXX[1]{{\textcolor{red}{\textbf{x\ #1\ x}}}}

\title{\large \bf A robust and simple method for filling in masked data in astronomical images}

\correspondingauthor{Pieter van Dokkum\\$^{\dagger}$\,\url{https://github.com/dokkum/maskfill}}

\author[0000-0002-8282-9888]{Pieter van Dokkum}
\affiliation{Astronomy Department, Yale University, 219 Prospect St,
New Haven, CT 06511, USA}

\author[0000-0002-7075-9931]{Imad Pasha}
\affiliation{Astronomy Department, Yale University, 219 Prospect St,
New Haven, CT 06511, USA}

\begin{abstract}

Astronomical images often have regions with
missing or unwanted information,
such as bad pixels, bad columns, cosmic rays, masked objects, or
residuals from imperfect model subtractions.
In certain situations it can be essential, or preferable,
to fill in these regions.
Most existing methods use low order interpolations for this task.
In this paper a method is described that uses
the full information that is contained in the pixels just outside masked
regions. These edge pixels are extrapolated inwards, using iterative median filtering.
This leads to a smoothly varying spatial resolution within the filled-in
regions, and ensures seamless transitions
between masked pixels and good pixels. Gaps in continuous, narrow features can
be reconstructed with high fidelity, even if they are large.
The method is implemented in \code{maskfill}, an open-source MIT licensed Python
package.$^{\dagger}$
%\footnote{\url{https://github.com/dokkum/maskfill}}
Its performance is illustrated with several examples, and compared to several
alternative interpolation schemes.

\end{abstract}

\keywords{
Direct imaging (387)  --- Astronomical techniques (1684) --- Astronomy data reduction (1861) --- Astronomy data analysis (1858)
}
\section{Introduction}

Image masking serves various purposes.
%It is universally
%employed in situations where specific pixels failed to reliably record the flux of astronomical objects.
Detector defects, such as hot pixels, bad pixels, or bad columns, result in predictable
locations where data cannot be trusted or are missing altogether.
Cosmic ray hits can occur anywhere on the detector, producing short trails of very bright pixels \citep{leach:79}. 
Both detector defects and cosmic rays are routinely masked
in the early stages of the data reduction process.

Another reason for masking is if certain objects are unwanted. An example is masking of bright stars
and galaxies in data that are searched for low surface brightness emission \citep{greco:18dwarfs,montes:18,danieli:19coma}.
A variation on
this theme is the masking of residuals after image subtraction. Image subtraction is routinely performed in transient
photometry \citep{kessler:15_short}, searches for faint or spatially-extended objects near bright ones \citep{marois:06,dokkum:mrf},
continuum correction of narrow band data \citep{james:04,garner:22,lokhorst:22}, and a wide range of other contexts. In all these applications,
regions where the subtraction is not satisfactory (such as the centers of bright stars) are typically masked, so they do not impact the subsequent analysis.

In many cases masked pixels do not need to be filled in, either because there are independent, redundant, observations of the same sky positions
\citep[e.g.,][]{fruchter:02}, or because subsequent image analysis steps can explicitly handle masked data, e.g., profile fitting softwares such as \code{GALFIT} \citep{galfit} and \code{pysersic} \citep{Pasha:2023}.

However, there are exceptions when mask filling (also known as inpainting)
is desirable or necessary. Multiple redundant exposures are not always available, for instance in time series data. Also,
detector defects and cosmic rays
are sharp features that are best identified before resampling the data; this is why reduction pipelines often
remove these features from individual frames, even if multiple exposures are available \citep{kelson:03,neill:23}. 
Turning to masked stars and galaxies, convolutions such as Gaussian smoothing produce artifacts on both sides of sharp boundaries,
and these can be greatly reduced when the masks are (temporarily) filled in beforehand. Another reason to fill in masks  is to
properly account for missing flux when measuring intracluster light \citep{montes:18} or Galactic cirrus emission \citep{liu:23}. A recently developed
application is to obtain the best-possible local background for photometry; \citet{saydjari:22} show that inpainting
can lead to significant improvements in photometric accuracy
when dealing with spatially-varying backgrounds.
More generally, it can simplify the analysis, particularly
when doing object detection and characterization in large datasets.

Methods for filling in masked data range from the extremely simple (replacing all masked pixels by zeros, or a block-median of surrounding pixels) to the highly complex.
Examples in the latter category are various applications of
machine learning \citep[e.g.,][]{zhang:20,lomeli:22}, untrained convolutional networks
\citep{ulyanov:20}, and Gaussian process regression \citep{saydjari:22}.
Most widely-applied methods
use some form of direct interpolation, either in the form of low order 2D functions that are fit to the surrounding pixels
\citep{sakurai:01,popowicz:13,popowicz:15}, or by applying a
median filter with a size that exceeds that of the masked features \citep{kokaram:95,dokkumc:01,huang:02}.
An alternative direct approach, particularly suited to large scale defects, is to predict missing data by analyzing
the Fourier transform of the image \citep{cooray:20}.

In this paper a new mask-filling, or inpainting, method is presented that can be described as ``interpolation by extrapolation'': rather than
interpolating over a masked region, its edge pixels are extrapolated inward. Although very different in execution, it is similar
in spirit to the Fourier methods of \citet{cooray:20}. In its iterative implementation it is akin to
classic flood fill algorithms \citep{newman:79}, as well as the  cosmic ray identification code
\code{l.a.cosmic} \citep{dokkumc:01}.
The concept is introduced in \S\,2 and its
implementation in the \code{maskfill} Python script is described in \S\,3. Some examples of its operation are presented
in \S\,4. Comparisons to other interpolative algorithms and performance characterization is presented in \S\,5, and a short conclusion is presented in \S\,6.

\section{Methodology}
\subsection{Concept}

\begin{figure}[htbp]
  \begin{center}
  \includegraphics[width=1.0\linewidth]{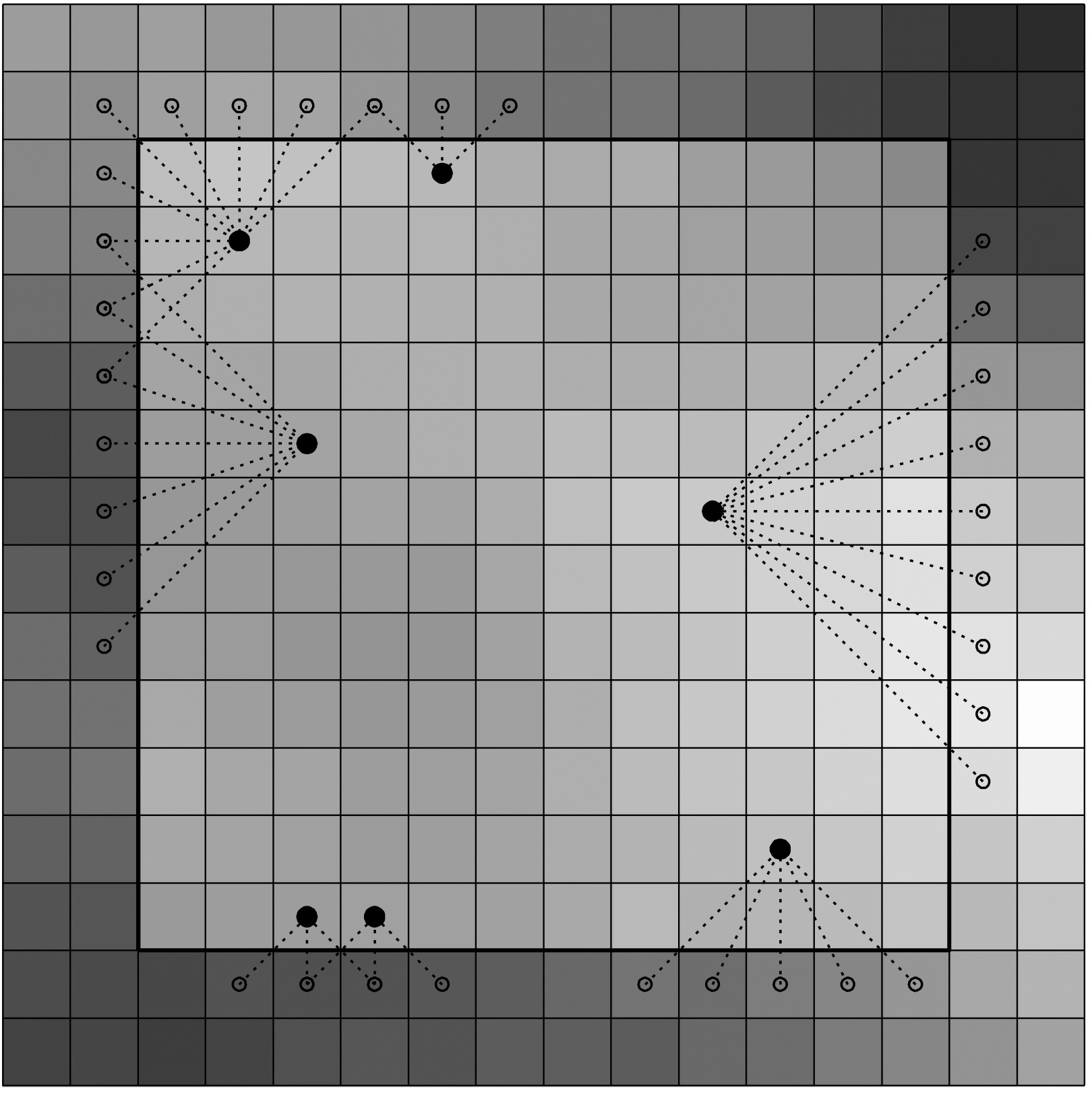}
  \end{center}
\vspace{-0.2cm}
    \caption{
Illustration of the concept. A small ($16\times 16$) section of an actual image is
shown.
The central $12\times 12$ pixels are shown behind a hypothetical semi-transparent mask.
Typically all masked pixels are replaced by the same value, or by a low order 2D surface.
However, it is clear that there are strong correlations between pixels just outside and
just inside the mask. These correlations are maintained if pixels near the edge
are filled by the mean or median of only the immediately-adjacent
edge pixels. Pixels that are closer to the center are filled by 
the mean of a larger number of edge pixels, with the number of contributing
pixels increasing with the distance to the nearest edge.
}
\label{method.fig}
\end{figure}

The central idea is that masked pixels that are adjacent to unmasked pixels should be treated in a
different way than
masked pixels that are far from any unmasked pixels. In the former case, the true values of the
masked pixels can reasonably be expected to be similar to those of their immediate neighbors, whereas in the latter
case there is much less information and a greater degree of smoothing is appropriate.
For a given pixel at location $(x,y)$ within a masked region, at a distance $d$
from the nearest edge, this concept can be expressed as
\begin{equation}
F_{x,y} = \frac{1}{2d+1}\sum_{n=-d}^{d} F_{x+d,y+n},
\label{method.eq}
\end{equation}
if the nearest edge is in the $+x$ direction and the pixel is not near a corner.
A graphic illustration of the concept is shown in Fig.\ \ref{method.fig}.

The number of edge
pixels that contribute to the value of
a masked pixel is $(2d+1)$.
Pixels near the center are averages of many edge pixels, whereas pixels close
to the edge are determined by their immediate neighbors only. This implies
a spatially-dependent smoothing, maintaining high resolution information
near the edge and smoothing by $\sim\,$half the size of the 
masked region near the center.
\begin{figure*}[htbp]
  \begin{center}
  \includegraphics[width=0.65\linewidth]{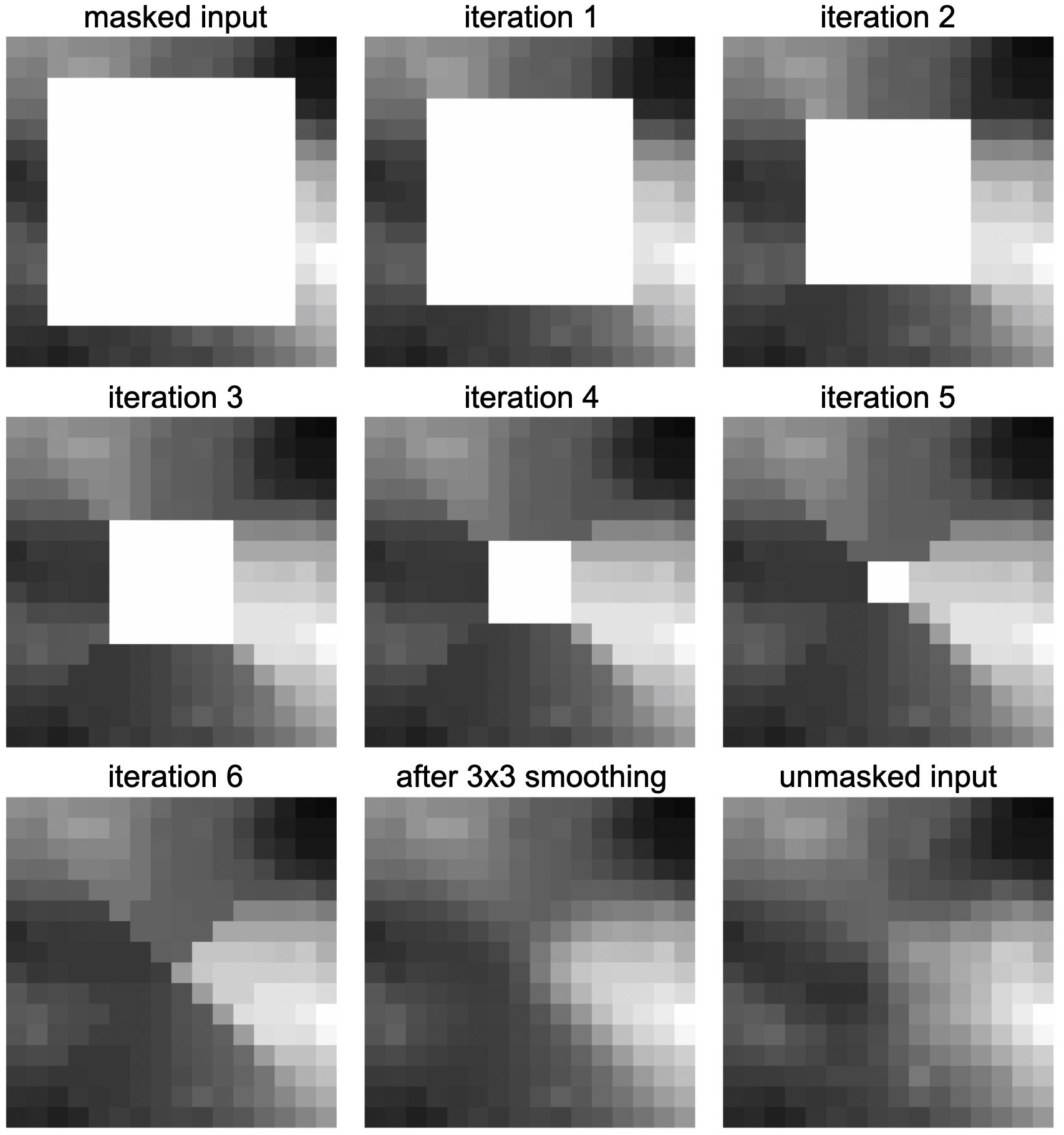}
  \end{center}
\vspace{-0.2cm}
    \caption{
Demonstration of the iterative extrapolation of edge pixels. The masked input image
is shown at upper left. In each iteration, pixels inside the mask are replaced by
the median of their closest neighbors, by applying a $3\times 3$ filter. The final
step is to apply a $3\times 3$ boxcar filter to the now filled-in masked region. In this
case, the filled-in
region is quite similar to the actual data, shown in the lower right panel.
}
\label{demo.fig}
\end{figure*}
\subsection{Iterative Extrapolation of Edges}
While possible, it is not straightforward to turn Eq.\ \ref{method.eq} directly
into a practical and efficient application.
It would be necessary to find all connected masked pixels and their edges, a task that is complicated by
the fact that, in practice, masked regions overlap. As an example, bad columns typically
intersect many cosmic rays, producing complex shapes.

Fortunately we can simplify the problem, by making use of the hierarchy that is inherent
in Eq.\ \ref{method.eq}. Each pixel that is adjacent to an edge is the average of its
three immediate neighbors. As can be deduced from the dotted lines in Fig.\ \ref{method.fig},
{\em this same rule  applies
to pixels that are far away from an edge} -- the only difference is that the values of the
three adjacent pixels are not known a priori. The solution, then, is to iteratively extrapolate the edges
inward, so that each successive layer can be determined from the previous one, using the same
operation.

Successive steps in this process are shown in Fig.\ \ref{demo.fig}, created with the
default version of \code{maskfill} (described below). In each iteration, a $3\times 3$
median filter is applied to the image (a mean filter produces similar results), until
all masked pixels are filled in. The final step in the process is
the application of a $3\times 3$ boxcar filter to all
pixels within the masked region, to smooth out sharp ridges.
These ridges are evident after the sixth iteration, and an artifact of the application
of Eq.\ 1 to a square region. The extrapolations of the four sides are largely
independent until they meet in the middle, producing an X-pattern that is characteristic
of the method.

The reconstructed image is an excellent match to the truth,
shown in the bottom right panel of Fig.\ \ref{demo.fig}. This is not always the case
(see \S\,\ref{conclude.sec}), but it is not unusual: the match is generally good
when the mask interrupts a continuous feature, such as the emission going from top left
to bottom right in this example.
\begin{figure*}[htbp]
  \begin{center}
  \includegraphics[width=0.9\linewidth]{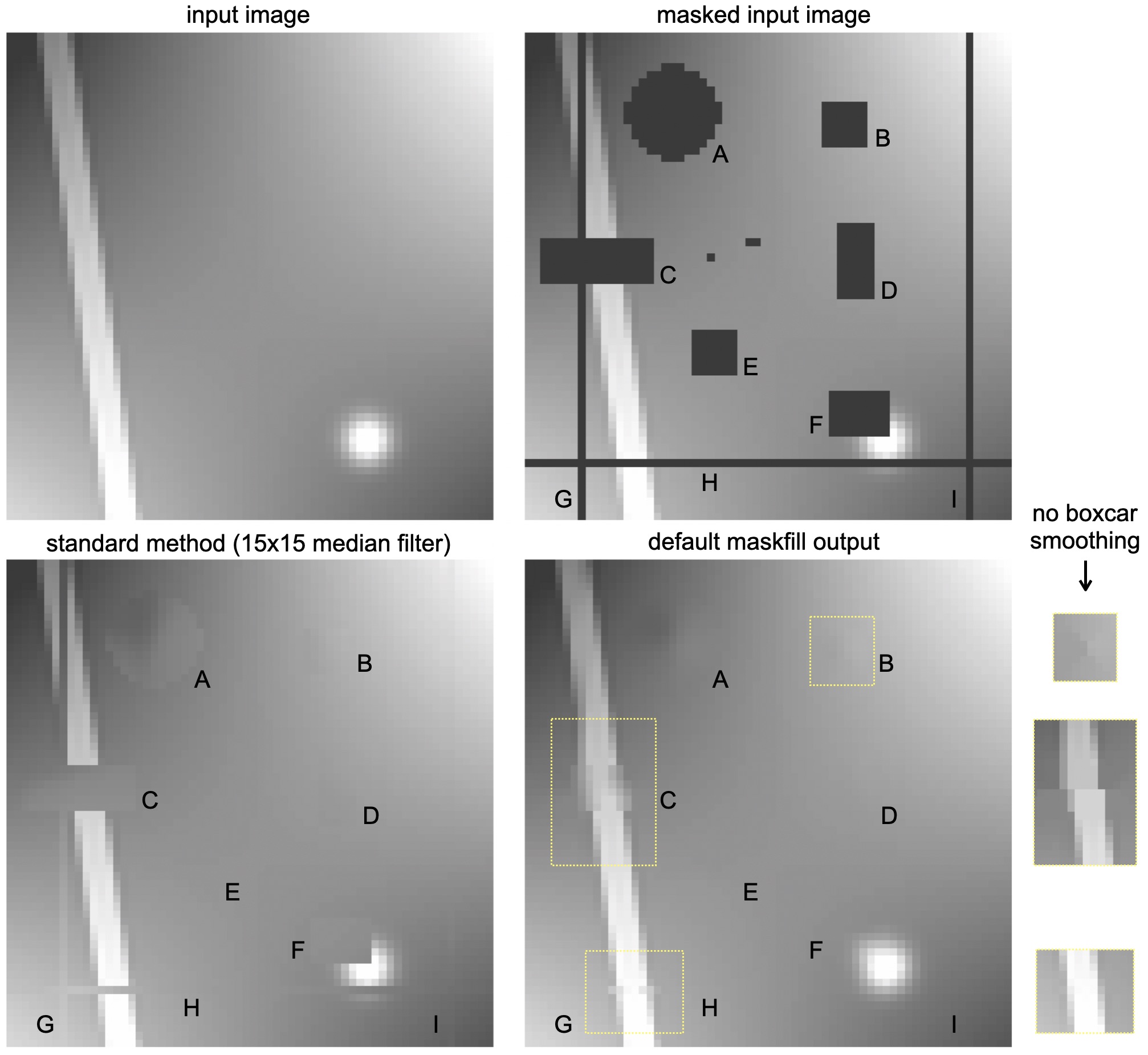}
  \end{center}
\vspace{-0.2cm}
    \caption{
Application to a synthetic image with a variety of structures and masks.
The input image and the masked input image are shown at the top.
A standard method for filling in masked regions, using a $15\times 15$ pixel median filter
applied to unmasked pixels, is shown at lower left. The lower right panel shows the \code{maskfill} output.
The masked row and columns (G, H, I) are nearly identical to the input image, and both
the star (F) and the linear feature (C) are reconstructed reasonably well. Insets at right
show the effects of setting the \code{--nosmooth} flag in \code{maskfill}. 
}
\label{gradient.fig}
\end{figure*}

\section{Implementation}

The method is implemented in the Python package \code{maskfill}, distributed via
\code{github} ({\url{https://github.com/dokkum/maskfill}}) or the Python Package Index (\code{PyPI}). 
Given an input image and a mask, the code 
\begin{enumerate}
    \item identifies masked pixels that border at least one non-masked pixel via a 2D convolution with a $3\times 3$ uniform kernel and sum operation;
    \item processes all identified pixels, filling with the median (or mean) of the adjacent non-masked pixels, depending on the chosen window size;
    \item updates these pixels in the output image, then repeats, using the new image as the base, until all pixels are filled; and 
    \item (optionally) performs a final $3\times3$ boxcar smoothing within the filled regions.
\end{enumerate}

Note that the median filter is not a \textit{running} median, in which an updated pixel may affect the value of the next-processed pixels. 
%In each iteration, all possibly-fillable pixels are determined only using the extant pixels prior to that iteration. 
A $3\times 3$ filter and median filtering will
in almost all cases produce the best results, and the code is generally fast enough that selecting a larger window or mean instead of median are not necessary. Increasing the window size can help in the special case where the masks are too small, such that
some of the edge pixels are defects or have other undesired values.

The code can be run either from the command line or from within Python. From the command line, one can (at simplest) run 
\begin{verbatim}
> maskfill in.fits mask.fits out.fits
\end{verbatim}
where \code{in.fits} is the input image, \code{mask.fits} is the mask containing 0s and 1s, with 1s indicating masked pixels, and \code{out.fits} the file to which to save the filled output. Optional parameters include the nature of the filter (mean or median, with median
the default), the size of the filter (default 3), a flag to omit the final smoothing step if desired, and a flag to write out a fits image after each iteration. If smoothing is enabled (default), another extension in the fits file will contain the unsmoothed version for comparison.\\
Within Python, one can call 
\begin{verbatim}
    from maskfill import maskfill 
    out,_ = maskfill('in.fits','mask.fits')
\end{verbatim}
More details about the usage can be found in the online documentation; the package has other convenience features not described here.

\section{Examples}
Here we describe several use cases and examples of the \code{maskfill} package applied to different images for different inpainting purposes. 

\subsection{Synthetic Image}
The method is first applied to
an artificial image of $64\times 64$ pixels, with a smoothly varying background, a linear feature, and a star.
Regions of varying shapes and sizes are masked, including a column, a row, and several
rectangles (see Fig.\ \ref{gradient.fig}).
Results from a standard method to fill in the mask are shown in the bottom left panel of
Fig.\ \ref{gradient.fig}. Here a $15\times 15$ pixel median filter was applied to the image, excluding
all masked pixels. The size of the filter just exceeds the dimensions of the largest masked region.
As expected, all masked regions, including those that intersect the linear feature
(C) and the star (F), are filled with the local background.
This fill method works well for regions that are far away from objects: B, D, E, and
portions of the masked row (H) and columns (G, I). 

The output from \code{maskfill} is shown in the lower right panel. The code performs equally well
as the simple $15\times 15$ median filter in empty regions, and much better where the masks intersect
objects. The masked portions of the linear feature (C) and the star (F) are reconstructed fairly well.
The reconstructions of the masked row and columns are nearly perfect; for features with a width of
1 or 2 pixels, such as bad pixels, hot pixels, bad columns, and many cosmic rays,
the code reduces to a straightforward (and optimal) interpolation of the locally-adjacent pixels.

As noted above, the default settings of \code{maskfill} include a $3\times 3$ boxcar smoothing at the end. The
insets at right show the effects of turning this smoothing off (with the \code{--nosmooth} flag).
The reconstruction of the sharp linear feature is improved, at the cost of having a faint
X-pattern in B.

\begin{figure}[htbp]
  \begin{center}
  \includegraphics[width=1.0\linewidth]{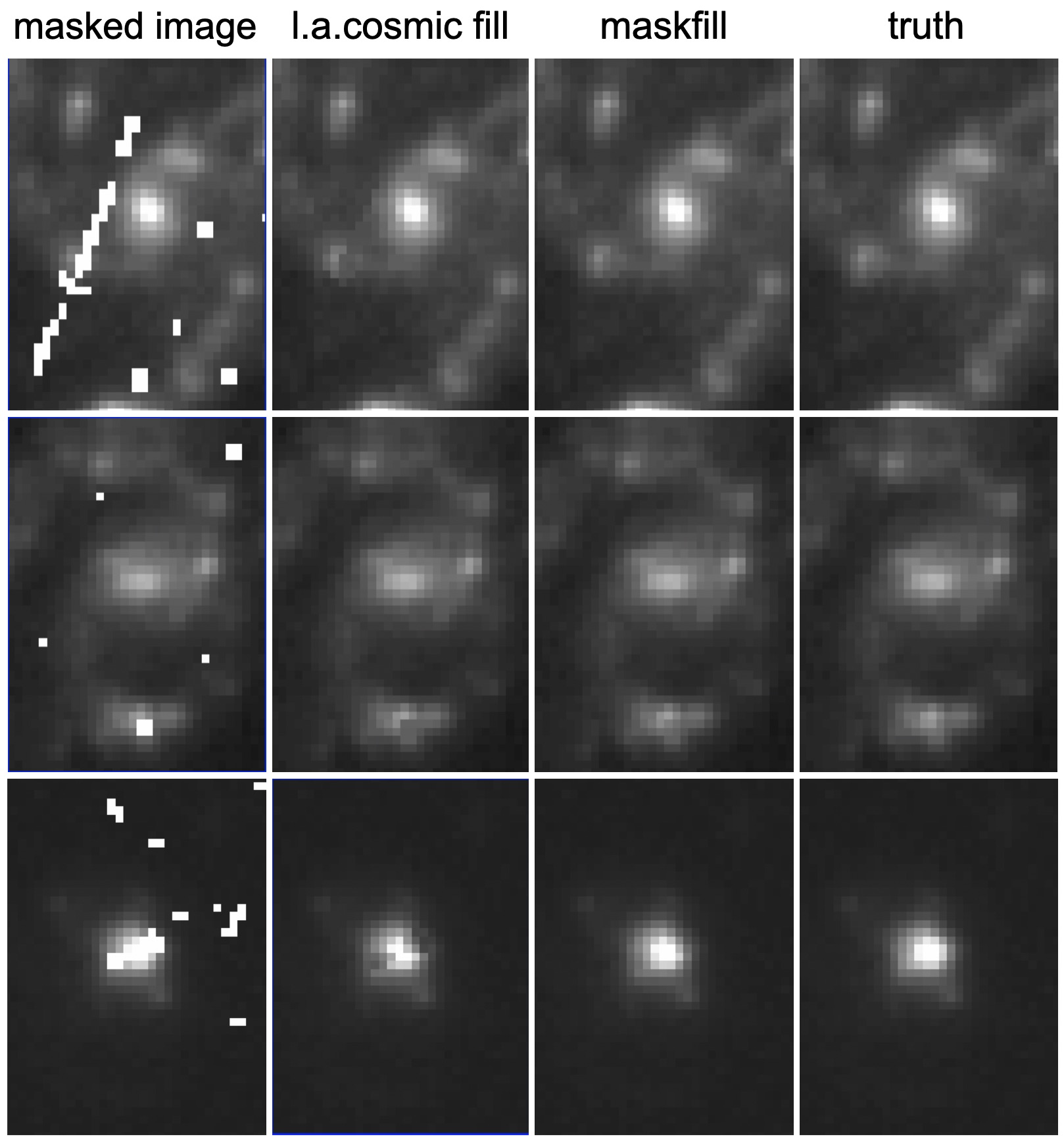}
  \end{center}
\vspace{-0.2cm}
    \caption{
Replacement of
cosmic rays, in $1\farcs 2\times 1\farcs 7$ sections of a deep HST image. The
cosmic rays were obtained from an independent image and added to the drizzled data.
The \code{maskfill} code performs slightly better than the fill routine
in \code{l.a.cosmic} \citep{dokkumc:01}, although differences are small.
}
\label{cosmics.fig}
\end{figure}

\begin{figure*}[htbp]
  \begin{center}
  \includegraphics[width=0.9\linewidth]{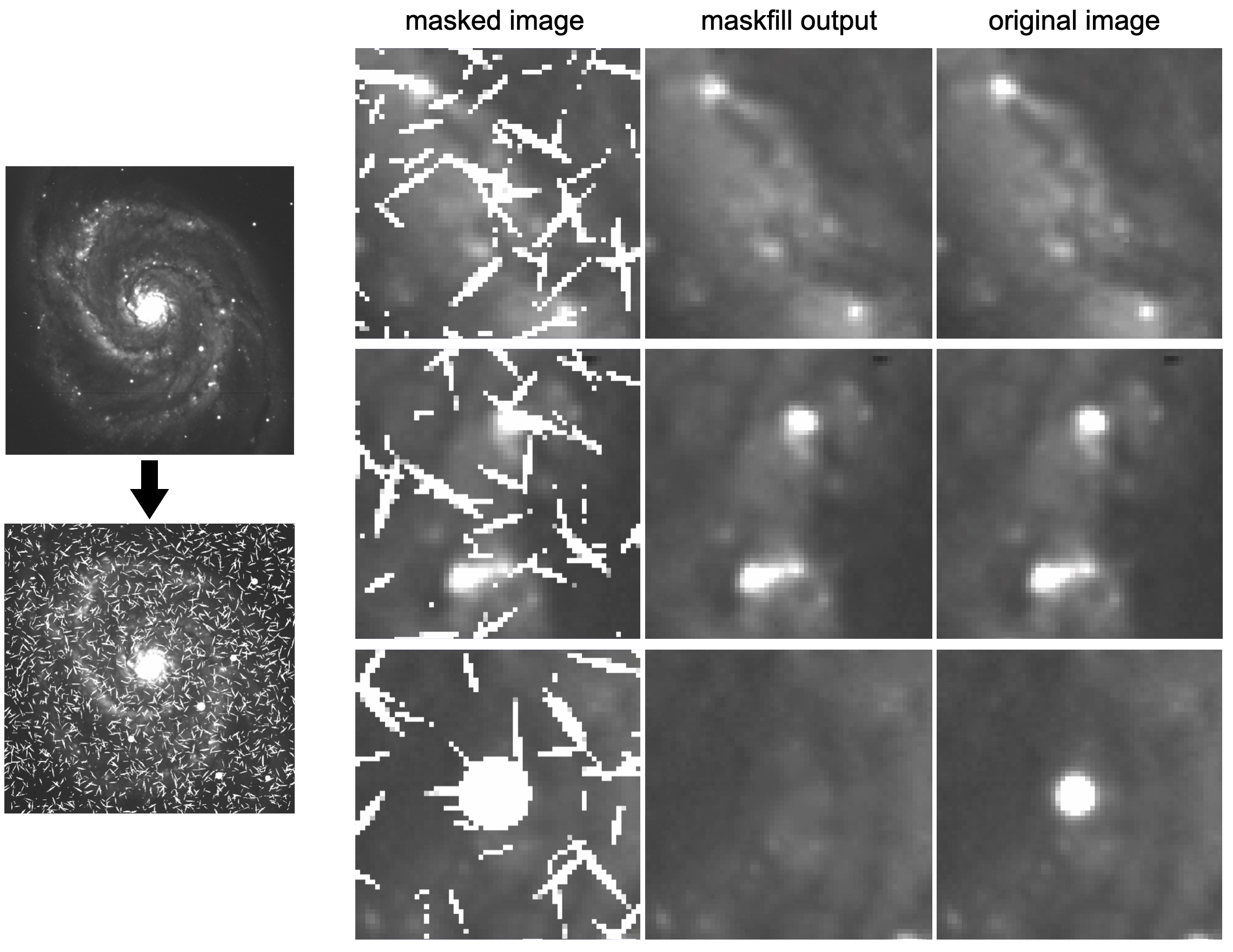}
  \end{center}
\vspace{-0.2cm}
    \caption{
Demonstration on the \code{IRAF} M51 image, with a mask containing
3000 thick cosmic ray-like objects. The \code{maskfill} reconstructions, with default
parameters, are
shown in the middle column, with the original images at right.
The bottom row shows a masked star.
}
\label{m51.fig}
\end{figure*}

\subsection{Cosmic Ray Replacement}

While \code{maskfill} cannot be used to {\em detect} cosmic rays, it is well-suited to replacing
them with plausible values after they are found. A test image was created by adding actual
cosmic rays (identified in an HST WFPC2 image) to a deep HST UVIS image (from program GO-17301). 
Three example galaxies are shown in Fig.\ \ref{cosmics.fig}.
The performance of \code{maskfill} is compared to that of \code{l.a.cosmic} \citep{dokkumc:01}.
After identifying cosmic rays, \code{l.a.cosmic} produces a cleaned image by replacing
masked pixels with the median of neighboring pixels, using a fixed $5\times 5$ filter.
This size is adequate for cosmic rays, as they are always narrow in at least one dimension.

The performance of \code{maskfill} is indistinguishable from that of the fill step of \code{l.a.cosmic}
in the vast majority of cases, with both methods producing excellent results. In the rare cases where there is
a difference, \code{maskfill} performs better. An example is shown in the bottom panels of Fig.\ \ref{cosmics.fig}:
a cosmic ray covers the center of a compact galaxy, and the $5\times 5$ filter of \code{l.a.cosmic} is
too large to reconstruct the affected pixels.
For optimal results, \code{maskfill} can simply be run after \code{l.a.cosmic}, using
the original image and the mask that \code{l.a.cosmic} produces as inputs.

\begin{figure*}
    \centering
    \includegraphics[width=0.88\linewidth]{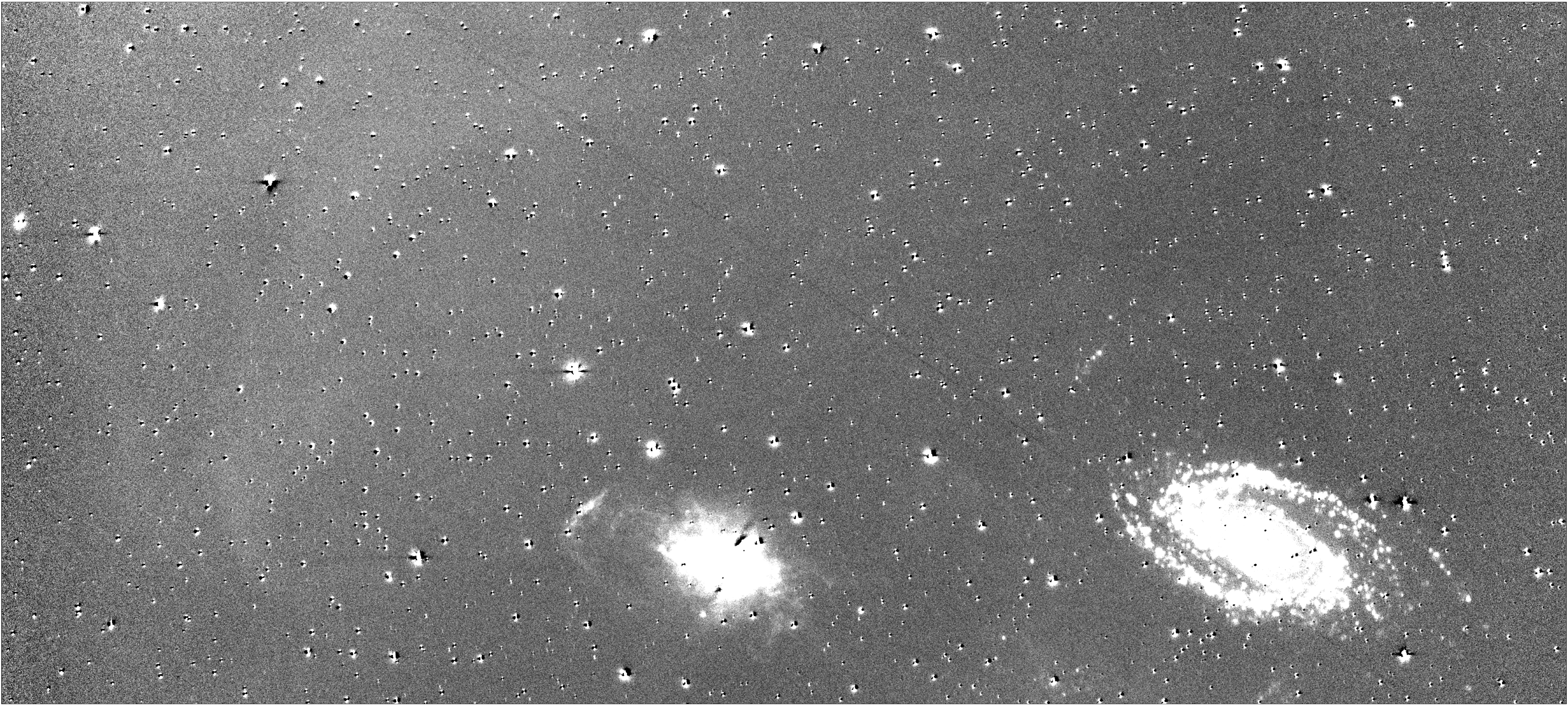}
    \includegraphics[width=0.88\linewidth]{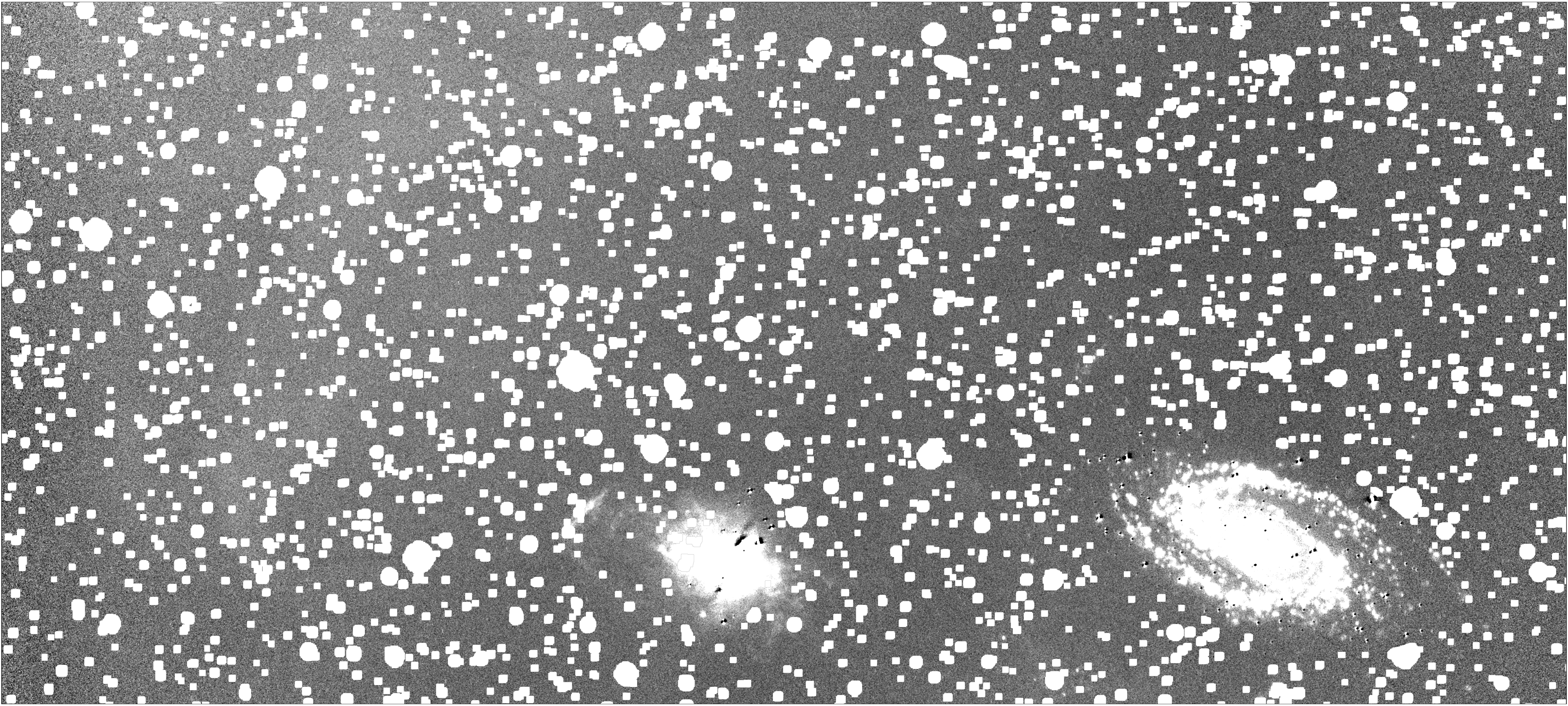}
    \includegraphics[width=0.88\linewidth]{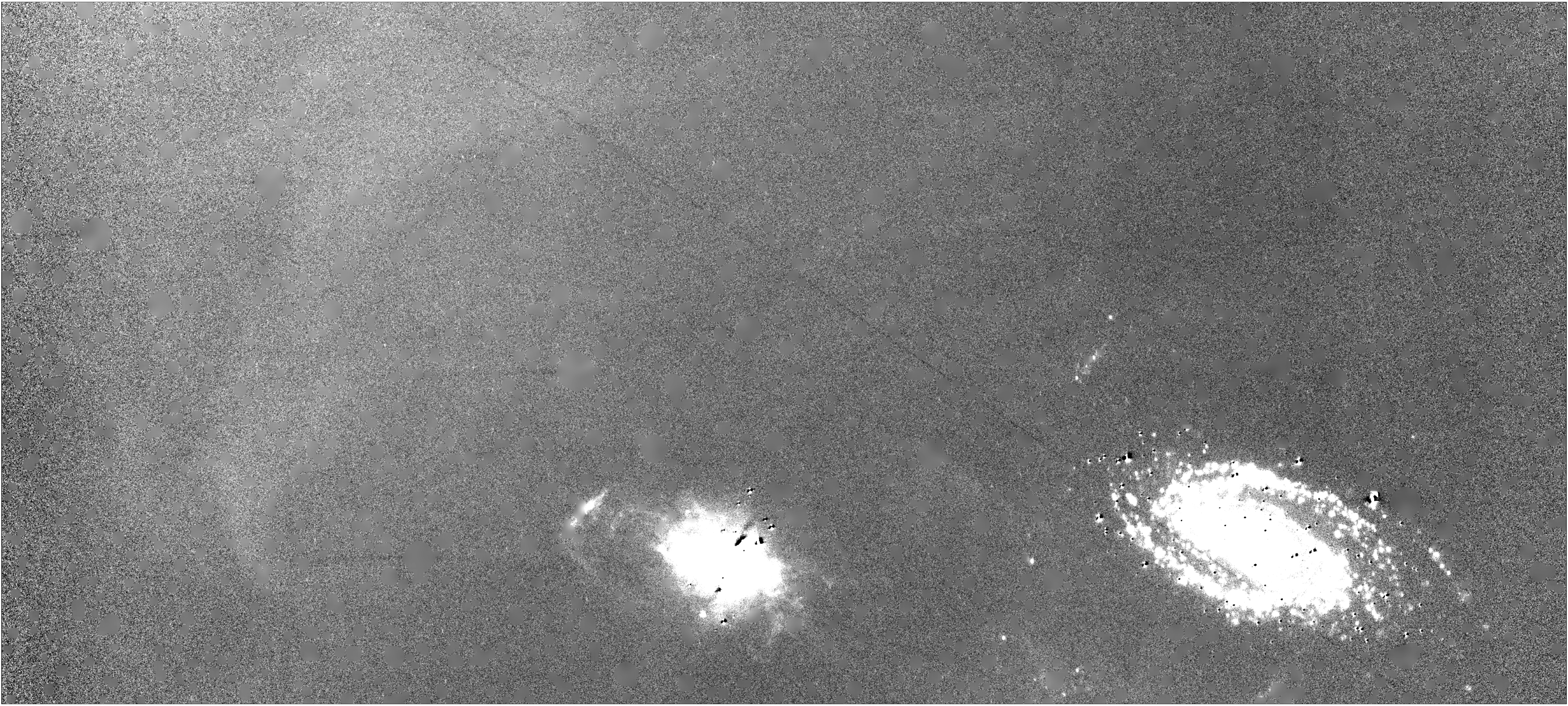}
    \caption{Cutout of a Wide-field narrowband $H\alpha$ image of M81 and M82 obtained with the Dragonfly Spectral Line Mapper pathfinder instrument\citep{Pasha:2021,Lokhorst:2022}. The size of this image is (3600, 2500). \textit{Top}: Image after continuum subtraction, with stellar residuals due to saturation and differences in the PSFs. \textit{Center}: Masks for the residual star locations. Individual masked objects often cover $\sim 10^2$ pixels. \textit{Bottom}: \code{maskfill} output, highlighting the way large scale features can be more easily identified. Note that due to the significant size of each mask (in pixels), \code{maskfill} was run with a window size of 11. Despite the size and number of masks, \code{maskfill} completed in $\sim$30\,s.}
    \label{fig:m81}
\end{figure*}

\subsection{Sections of M51}

For a more challenging test, the iconic {\code IRAF} $512\times 512$ pixel image of M51 was degraded by adding
3000 wide and long cosmic ray-like objects. A few bright stars in the image were also masked. 
In Fig.\ \ref{m51.fig} the \code{maskfill} reconstructions of several regions are compared to the
unmasked input. The filled images are very similar to the original images, even in cases where
several masks intersect.
The bottom row of Fig.\ \ref{m51.fig} shows a masked star. The filled-in mask is a plausible
continuation of the surrounding emission.

\subsection{Wide Field Low Surface Brightness Imaging}

The detection of
low surface brightness (LSB) features, such as tidal tails and diffuse galaxies, can
be challenging,  particularly in crowded fields. In Figure \ref{fig:m81} we show cutouts of a wide-field continuum-subtracted $H\alpha$ image of the M81 group, taken with the Dragonfly Spectral Line Mapper pathfinder \citep{Pasha:2021,Lokhorst:2022}. A typical contaminant in continuum-subtracted imaging are residuals of bright stars, as saturation, color terms, and imperfect PSF matching can lead to strong residuals. These residuals are masked in the middle panel. The image size ($3600 \times 2500$ pixels) and the size of each mask ($\sim 10^2$ pixels) make this a taxing example of a \code{maskfill}'s  use case. 

The bottom panel shows that the code performs well, producing a very clean image that makes large, nearly degree-scale low surface brightness features easy to see by eye.
In this example the window size was set to 11 instead of the default 3 pixels,
as the masks do not always cover the residuals completely.
The runtime was $\sim$30\,s. We discuss the overall runtime performance of \code{maskfill} further in Section \ref{comparisons}; future work aims to bring it
down further via the use of optimized just-in-time compilation packages.

\begin{figure*}
    \centering
    \includegraphics[width=0.99\linewidth]{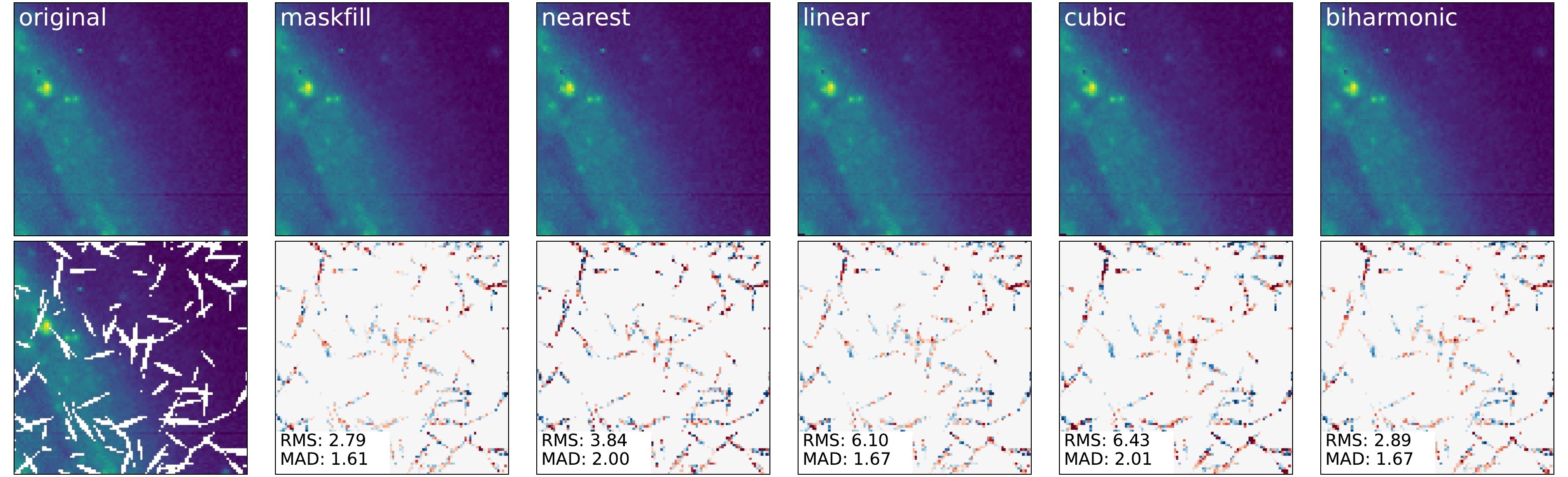} \\
    \vspace{0.2cm}
    \includegraphics[width=\linewidth]{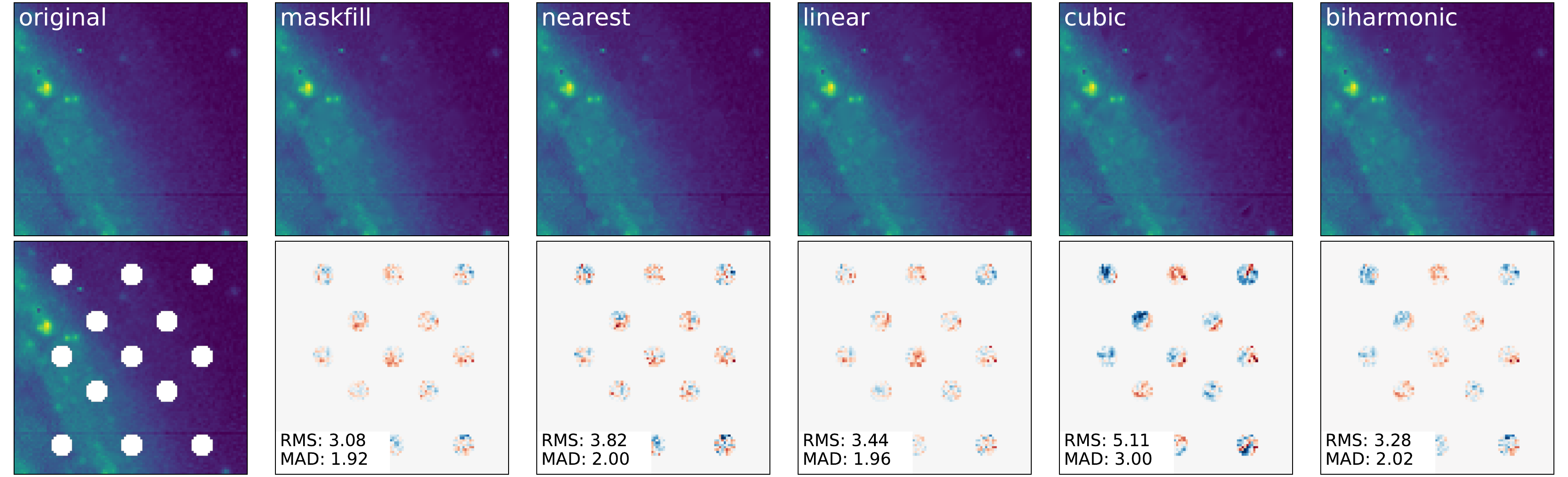} \\
    \vspace{0.2cm}
    \includegraphics[width=\linewidth]{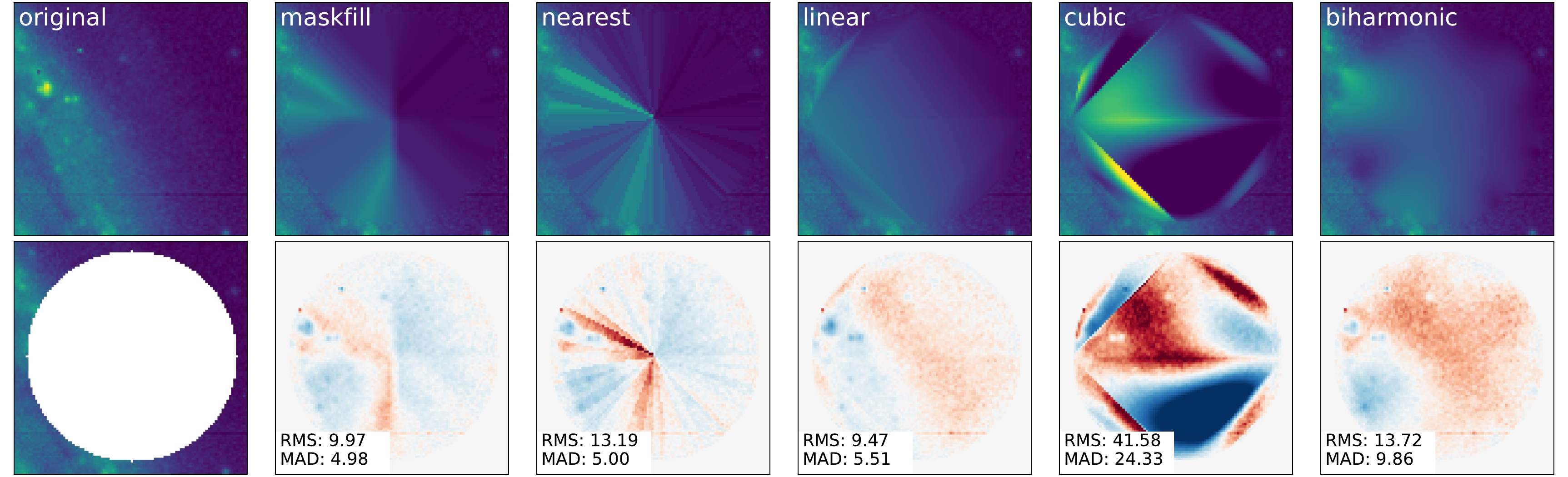}
        \caption{Comparison of different interpolation schemes. \textit{Top}: Interpolations in the presence of many small masks (e.g., the case of cosmic rays). We show the original image (masked version below), followed by the outputs from \code{maskfill}, nearest-neighbor, as well as linear, cubic, and biharmonic splines. The bottom panels show the residuals normalized to the original image, and the RMS and MAD are shown. \textit{Middle}: Same as top panels, but for 13 small masks (e.g., the star-like case). \textit{Bottom}: A single, huge mask. In this case, \code{maskfill} was run with a window size of 5 pixels. Overall, we find that \code{maskfill} outperforms or matches the other methods presented in all cases, and is particularly more robust for large masks. The metrics in this figure can also be found in Table \ref{tab:scores}.}
    \label{fig:method_comparison}
\end{figure*}

\begin{figure*}
    \centering
    \includegraphics[width=\linewidth]{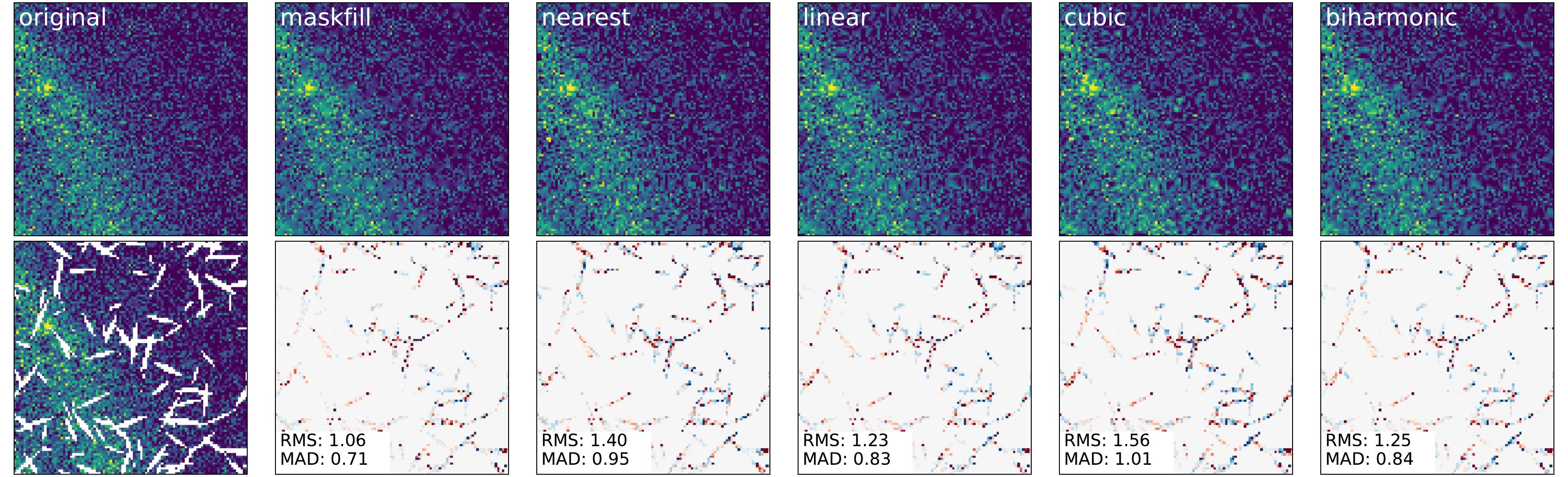} \\
    \vspace{0.2cm}
    \includegraphics[width=\linewidth]{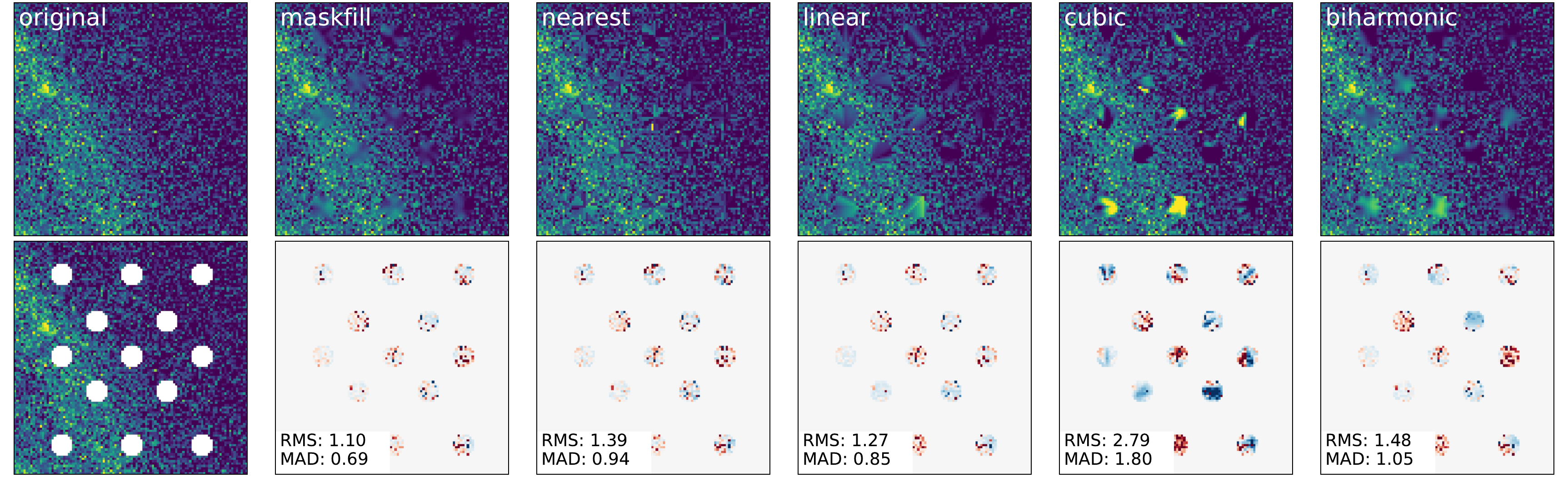} \\
    \vspace{0.2cm}
    \includegraphics[width=\linewidth]{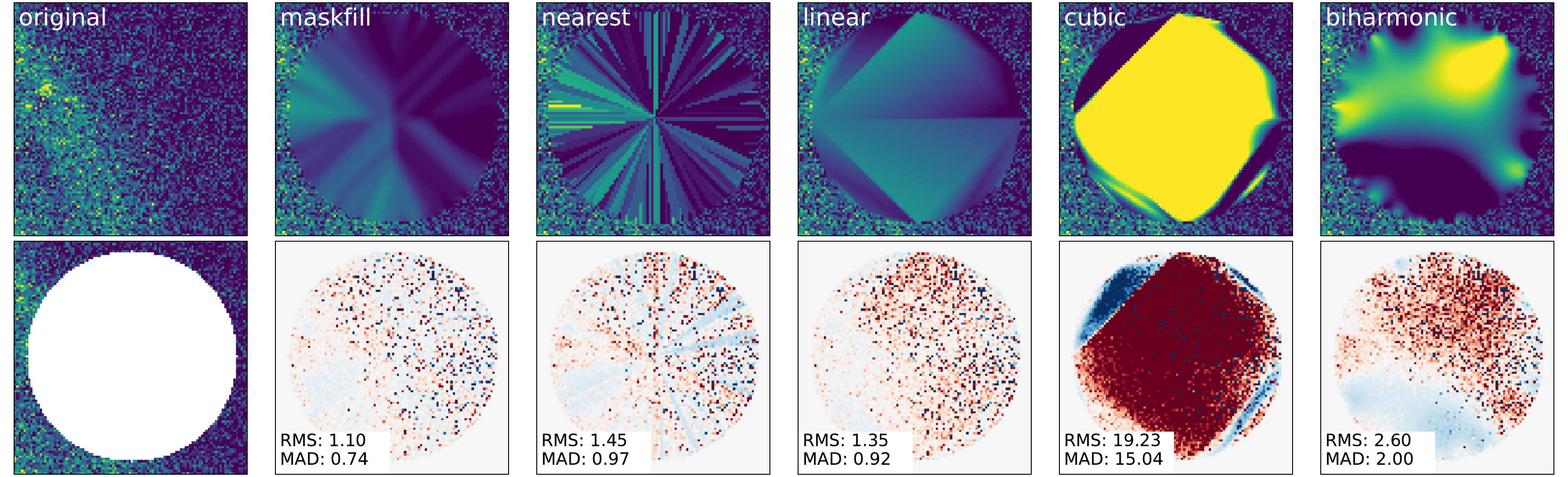}
    \caption{Interpolation schemes in the presence of noise, with the same mask examples presented in Figure \ref{comparisons}. Here, we add Gaussian noise to the M51 image with a scale $\sim$the median pixel value, simulating a low signal-to-noise image. The quoted RMS and MAD have been normalized to the scale of the added noise for easier comparison to the noiseless test. We find that in the presence of noise, \code{maskfill} performs significantly better than the other interpolation methods both in RMS and MAD, as well as visually (particularly when the mask gets large). }
    \label{noise_comparison}
\end{figure*}

\begin{deluxetable*}{lcccccc}[htp]
\label{tab:scores}
\tablecaption{Interpolation Performance Comparison for the three test mask cases and two noise scenarios.}
\tablehead{
    \colhead{Method} &
    \colhead{CRs} &
    \colhead{CRs + Noise} &
    \colhead{Star Masks} &
    \colhead{Star Masks + Noise} &
    \colhead{Large Mask} &
    \colhead{Large Mask + Noise}}
\startdata
Maskfill & 2.79 (1.61) & 1.06 (0.71) & 3.08 (1.92) & 1.10 (0.69) &9.97 (4.98)&1.10 (0.74)  \\
Nearest Neighbor & 3.84 (2.00) & 1.40 (0.95) & 3.82 (2.00) & 1.39 (0.94) &13.2 (5.00)&1.45 (0.97)  \\
Linear & 6.10 (1.67) & 1.23 (0.83) & 3.44 (1.96) & 1.27 (0.85) &9.47 (5.51)&1.35 (0.92)  \\
Cubic & 6.43 (2.01) & 1.56 (1.01) & 5.11 (3.00) & 2.79 (1.80) &41.6 (24.3)& 19.2 (15.0)  \\
Biharmonic Spline & 2.89 (1.67) & 1.25 (0.84) & 3.28 (2.02) & 1.48 (1.05) &13.7 (9.86)&2.60 (2.00)  \\
\enddata
\tablecomments{Values are root mean square errors (RMSE), with the median absolute deviation (MAD) in brackets. The nearest-neighbor, linear, and cubic routines are from the \code{scipy.interpolate} package, while the biharmonic spline is from the \code{scikit-image} package.}
\end{deluxetable*}

\vspace{-0.3cm}
\section{Comparisons and Performance}\label{comparisons}

Here we assess the filling quality of \code{maskfill} compared to other standard interpolative routines, including nearest-neighbor, linear, cubic, and biharmonic interpolation.

First, we assess the root mean square error (RMSE) and median absolute deviation (MAD) between an un-destructed image and a masked image across the various methods, taking three reasonably separated cases: the ``cosmic ray case'', in which there are many small masks of $\sim$few pixels each, the ``star'' or ``artifact'' case, in which there are several medium-sized circular masks, and the ``large mask'' case. We use the M51 image for this test, and to avoid the (expected) residuals from foreground stars and the center of the galaxy dominating the comparison, we choose an arbitrary, quiet region in the galaxy outskirts for this test. We conduct all tests both on the original image, which has very high signal-to-noise, as well as a version for which Gaussian noise has been added to artificially lower the signal-to-noise ratio.  

The results are presented in Figure \ref{fig:method_comparison}, Figure \ref{noise_comparison}, and Table \ref{tab:scores}. We find that \code{maskfill} generally has the lowest RMSE and MAD across these tests, and in particular, is robust against catastrophic fits in the case of noisy images. For the ``noiseless'' case (Figure \ref{fig:method_comparison}), \code{maskfill} has generally the best statistics, performing particularly well in the case of the very large mask. Several other methods, such as the linear and biharmonic spline, produce reasonably consistent residuals (though less so for the large mask). 

For the noise-added images, the differences become more stark --- \code{maskfill} has considerably lower residuals than the other methods. Particularly as the masks get larger, the other methods produce artifacts and bleed-through features which render the fills mostly unusable. Conversely, the medians-of-medians iterative routine used by maskfill is robust against the presence of edge pixels which are far from the center of the distribution.

\begin{figure*}
    \centering
     \includegraphics[width=\linewidth]{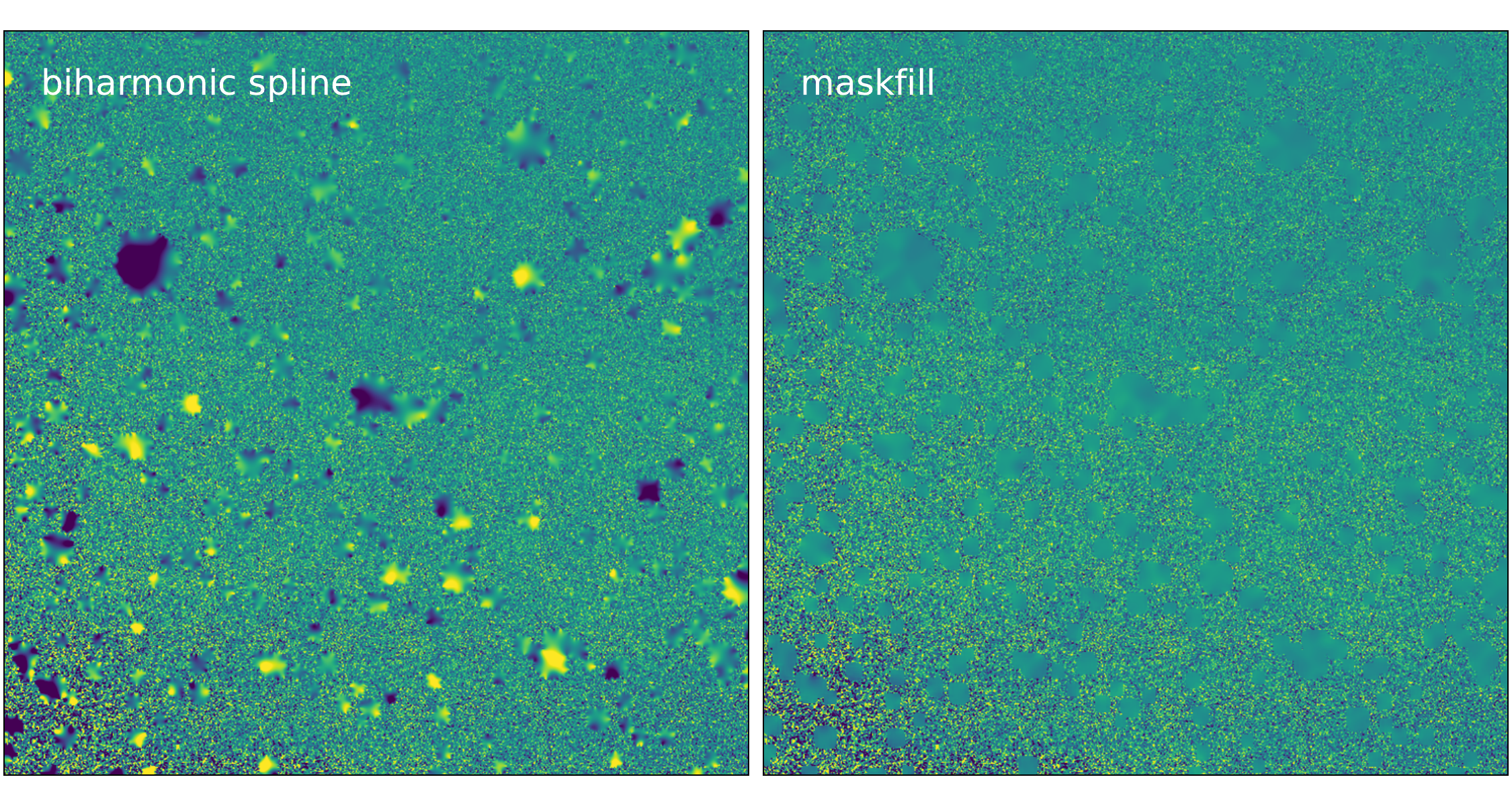}
    \caption{Comparison of the infilling between biharmonic spline (left) and \code{maskfill} (right) on a cutout of the Dragonfly M81 narrowband image. Aspects of the pixel distribution of the image appear to cause catastrophic outputs from the spline interpolation, while the \code{maskfill} algorithm is mostly unaffected. This is because a simple localized median effectively downweights outliers, and as the iterations work inward into the mask, further medians quickly damp out the effect of aberrant pixels near the mask edge. For example, the large mask in the upper left has strongly negative values in the biharmonic spline restoration due to strong ``bleed-through'' of a small set of discrepant pixels in the upper right hand corner just outside of the mask.}
    \label{m81_compare}
\end{figure*}
This point is emphasized further when examining real, low signal-to-noise data. In Figure \ref{m81_compare}, we show a $750\times 750$ cutout of the M81 image from Figure \ref{fig:m81}, comparing the biharmonic output (left) to that of \code{maskfill} (right). There are no tuning parameters for the biharmonic spline as implemented in \code{scikit-image}. In these situations, \code{maskfill} appears to be considerably more robust, with medians tamping down on outlier pixels and their effect limited to the track of several radial pixels into the mask (beyond which further medians tend to dampen them out). The poor inpainting from the spline can be reasonably explained by a combination of the two effects described above: the noise of the image, and the large size of the masks, both of which contribute to the inability of the algorithm to appropriately infill the masked pixels
without tuning. We note that much better results can be obtained with 
biharmonic splines that are tuned to the size of the largest masks \citep[see][]{popowicz:15}.

While it is beyond the scope of this paper to perform a comprehensive speed analysis across methods (particularly given the fact that there are different implementations of the other interpolations), we do carry out some simple tests of \code{maskfill}'s speed to inform use cases users might have. The runtime of the code scales directly not with the image size, but with the total number of masked pixels and the size of the average mask in the image. The former is because the code executes a loop over all (fillable) masked pixels in each iteration, i.e., the edges of all masked regions, and the latter determines the number of infilling iterations required for the algorithm to complete. The detection of fillable-masked pixels as the start of each iteration is a fast convolution which does not significantly impact the runtime; thus, in total, we expect the runtime to scale approximately linearly with the number of masked pixels.

\begin{figure}[htbp]
    \centering
    \includegraphics[width=\linewidth]{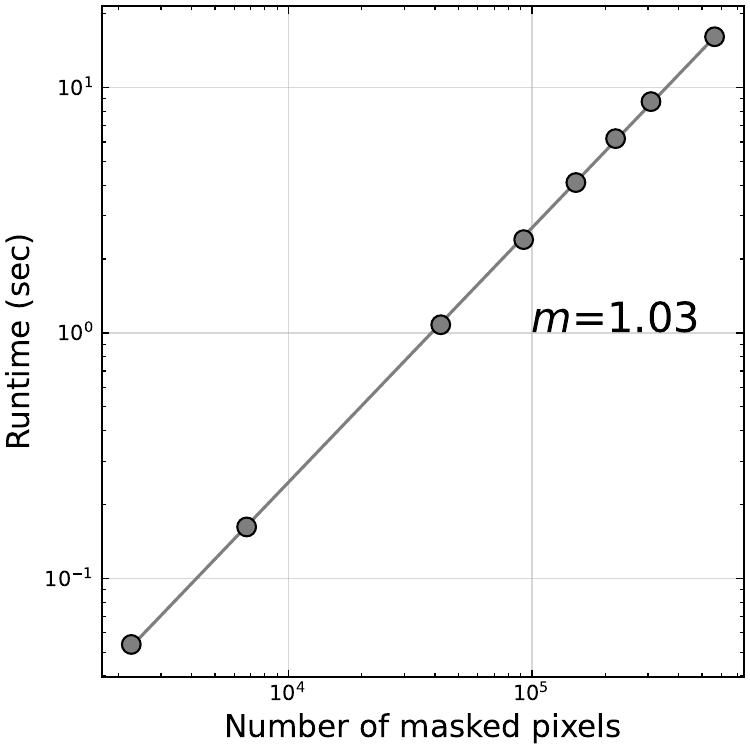}
    \caption{\code{Maskfill} runtime (in seconds) as a function of the number of masked pixels in the image, with a linear fit. In these examples the masks are distributed randomly, which means that the number of masked pixels scales directly with the size of the image. We find a fit coefficient of 1.03, demonstrating that the relationship is indeed linear. The normalization depends on the typical size of individual masks and the hardware that is used; for this particular example the fill speed is $\approx 5\times 10^4$\,pix\,sec$^{-1}$.}
    \label{speedtest}
\end{figure}

We assess the speed scaling of \code{maskfill} using the large M81 image and its associated mask, starting with a $100\times 100$ cutout and expanding sequentially to a $2000 \times 2000$ cutout. The masks in this image are generally uniformly sized and spaced, so progressively larger cutouts have more masked pixels. Figure \ref{speedtest} demonstrates that the runtime scales approximately linearly with number of masked pixels as expected; a fit to the runtimes gives $\log t_{\rm run} \propto 1.03 \log N_{\rm pix}$. 

The normalization depends on the number of iterations (and therefore
on the typical size of individual masks) as well as the hardware that is used.
In typical cases the fill speed will be $\sim 10^5$\,pix\,sec$^{-1}$ for small
masks such as bad pixels and cosmic rays, dropping to lower speeds for masked
stars and galaxies.

Because of the minimal dependencies and use of matrix-dominated operations, \code{maskfill} is well suited to current techniques for just-in-time compilation via, e.g., the \code{jax} or \code{numba} frameworks, potentially providing order-of-magnitude runtime improvements for larger images / larger masks; active development in this direction is underway.

% \begin{deluxetable*}{lcccc}
% \label{tab:speed}
% \tablecaption{Runtime Comparison}
% \tablehead{
%     \colhead{Method} &
%     \colhead{100x100 pix (STAR)} &
%     \colhead{100x100 pix (CR)} &
%     \colhead{512x512 pix (CR)} & 
%     \colhead{3600x2500 pix (M81)}
% }
% \startdata
% Maskfill & 15.2 ms $\pm$ 91.9 \textmu s & 36.2 ms $\pm$ 340 \textmu s & 942 ms $\pm$ 10.1 ms & 38.2 s \\
% Nearest Neighbor & 1.69 ms $\pm$ 31.6 \textmu s& 2.02 ms $\pm$ 22.6 \textmu s & 71.4 ms $\pm$ 525 \textmu s & 2.95 s $\pm$ 9 ms\\
% Linear &856 ms $\pm$ 7.28 ms & 682 ms $\pm$ 143 ms & 18.7 s $\pm$ 425 ms & \\
% Cubic & 754 ms $\pm$ 112 ms & 773 ms $\pm$ 15.8 ms & 18.7 s $\pm$ 825 ms & \\
% Biharmonic Spline & 1:28 $\pm$ 300 ms & 1:08 $\pm$ 361 ms & ** & **\\
% \enddata

% \tablecomments{In the implementation of the biharmonic spline from the \code{verde} package, we could not successfully infill the 512x512 image without the kernel crashing.}

% \end{deluxetable*}

\section{Conclusions}
\label{conclude.sec}

This paper describes \code{maskfill}, a new method for filling in masked emission in astronomical images.
The method is encapsulated in Eq.\ 1, and implemented as an iterative inward extrapolation of the edges
of masked regions. It is available as a Python package through
\code{github} and the Python package index (PyPI). Although \code{maskfill.py} has several optional
parameters, no tuning should generally be necessary: the code works on masks of arbitrary shapes and sizes, always
beginning with the pixels that are closest to the edge. 

Several caveats and limitations should be mentioned. Even though the algorithm is well-defined 
it is difficult to assign an uncertainty to the filled-in values. The formal per-pixel noise
decreases with $d$, the distance to the edge of the mask,
according to $\sigma_{\rm mask} = \sigma_{\rm org} \left(2d+1\right)^{-0.5}$, with $\sigma_{\rm org}$ the
per-pixel noise outside of the mask. However, the reliability of the reconstruction depends on the
likelihood of encountering features that do not extend beyond the mask boundaries. 
The bottom row of Fig.\ \ref{m51.fig} is a good example of this:
if the circular mask had been created to cover a defect, the
reconstruction would have entirely missed the star at that location.
Fortunately, most detector defects and cosmic rays
are on scales of a few pixels, where the code is reliable and its
behavior is predictable. Larger masks are often created to mask objects, and in those cases
the whole point is {\em not} to match reality but to replace the masked region with something that
fits in with its surroundings.

Another limitation is the characteristic cross pattern that arises on smoothly varying backgrounds,
particularly within square regions. This is inherent in the method: the extrapolation begins on
the four edges and then works its way inward. When the filled-in pixels finally meet in the center, they
each represent a different edge, with almost no knowledge of the other three edges. The boxcar smoothing
step at the end reduces this artifact, at a cost of a slight loss of resolution. This trade-off
is illustrated in Fig.\ \ref{gradient.fig}. 

That said, we find \code{maskfill} to be a relatively fast, intuitive, and robust method for mask infilling which relies only on the images at hand.

\begin{acknowledgements}
We thank the referee, Adam Popowicz, for insightful comments and for the suggestion
to compare the method explicitly to other interpolation schemes.
The {\texttt{maskfill}} code makes use of \href{http://www.numpy.org}{\code{NumPy}} \citep{numpy}; \href{http://www.scipy.org}{\code{SciPy}} \citep{scipy};
and \href{http://www.astropy.org}{\code{Astropy}} \citep{astropy2018}. Several example images were created with \href{https://iraf-community.github.io/}{\code{IRAF}},
the Image Reduction and Analysis Facility \citep{iraf1,iraf2}.
\end{acknowledgements}

\bibliographystyle{aasjournal}
\bibliography{master.bib}

\begin{thebibliography}{}
\expandafter\ifx\csname natexlab\endcsname\relax\def\natexlab#1{#1}\fi
\providecommand{\url}[1]{\href{#1}{#1}}
\providecommand{\dodoi}[1]{doi:~\href{http://doi.org/#1}{\nolinkurl{#1}}}
\providecommand{\doeprint}[1]{\href{http://ascl.net/#1}{\nolinkurl{http://ascl.net/#1}}}
\providecommand{\doarXiv}[1]{\href{https://arxiv.org/abs/#1}{\nolinkurl{https://arxiv.org/abs/#1}}}

\bibitem[{{Cooray} {et~al.}(2020){Cooray}, {Takeuchi}, {Yoda}, \&
  {Sorai}}]{cooray:20}
{Cooray}, S., {Takeuchi}, T.~T., {Yoda}, M., \& {Sorai}, K. 2020, \pasj, 72,
  61, \dodoi{10.1093/pasj/psaa038}

\bibitem[{{Danieli} \& {van Dokkum}(2019)}]{danieli:19coma}
{Danieli}, S., \& {van Dokkum}, P. 2019, \apj, 875, 155,
  \dodoi{10.3847/1538-4357/ab14f3}

\bibitem[{{Fruchter} \& {Hook}(2002)}]{fruchter:02}
{Fruchter}, A.~S., \& {Hook}, R.~N. 2002, \pasp, 114, 144,
  \dodoi{10.1086/338393}

\bibitem[{{Garner} {et~al.}(2022){Garner}, {Mihos}, {Harding}, {Watkins}, \&
  {McGaugh}}]{garner:22}
{Garner}, R., {Mihos}, J.~C., {Harding}, P., {Watkins}, A.~E., \& {McGaugh},
  S.~S. 2022, \apj, 941, 182, \dodoi{10.3847/1538-4357/aca27a}

\bibitem[{{Greco} {et~al.}(2018){Greco}, {Greene}, {Strauss}, {Macarthur},
  {Flowers}, {Goulding}, {Huang}, {Kim}, {Komiyama}, {Leauthaud}, {Leisman},
  {Lupton}, {Sif{\'o}n}, \& {Wang}}]{greco:18dwarfs}
{Greco}, J.~P., {Greene}, J.~E., {Strauss}, M.~A., {et~al.} 2018, \apj, 857,
  104, \dodoi{10.3847/1538-4357/aab842}

\bibitem[{Huang {et~al.}(2002)Huang, Cressie, \& Gabrosek}]{huang:02}
Huang, H.-C., Cressie, N., \& Gabrosek, J. 2002, Journal of Computational and
  Graphical Statistics, 11, 63.
\newblock \url{http://www.jstor.org/stable/1391128}

\bibitem[{{James} {et~al.}(2004){James}, {Shane}, {Beckman}, {Cardwell},
  {Collins}, {Etherton}, {de Jong}, {Fathi}, {Knapen}, {Peletier}, {Percival},
  {Pollacco}, {Seigar}, {Stedman}, \& {Steele}}]{james:04}
{James}, P.~A., {Shane}, N.~S., {Beckman}, J.~E., {et~al.} 2004, \aap, 414, 23,
  \dodoi{10.1051/0004-6361:20031568}

\bibitem[{{Jones} {et~al.}(2001){Jones}, {Oliphant}, {Peterson}, \& {et
  al.}}]{scipy}
{Jones}, E., {Oliphant}, T., {Peterson}, P., \& {et al.} 2001.
\newblock \url{http://www.scipy.org/}

\bibitem[{{Kelson}(2003)}]{kelson:03}
{Kelson}, D.~D. 2003, \pasp, 115, 688

\bibitem[{{Kessler} {et~al.}(2015){Kessler}, {Marriner}, {Childress},
  {Covarrubias}, {D'Andrea}, {Finley}, {Fischer}, \& {DES
  Collaboration}}]{kessler:15_short}
{Kessler}, R., {Marriner}, J., {Childress}, M., {et~al.} 2015, \aj, 150, 172,
  \dodoi{10.1088/0004-6256/150/6/172}

\bibitem[{Kokaram {et~al.}(1995)Kokaram, Morris, Fitzgerald, \&
  Rayner}]{kokaram:95}
Kokaram, A., Morris, R., Fitzgerald, W., \& Rayner, P. 1995, IEEE Transactions
  on Image Processing, 4, 1509, \dodoi{10.1109/83.469932}

\bibitem[{{Leach} \& {Gursky}(1979)}]{leach:79}
{Leach}, R.~W., \& {Gursky}, H. 1979, \pasp, 91, 855, \dodoi{10.1086/130599}

\bibitem[{{Liu} {et~al.}(2023){Liu}, {Abraham}, {Martin}, {Bowman}, {van
  Dokkum}, {Janssens}, {Chen}, {Keim}, {Lokhorst}, {Pasha}, {Shen}, \&
  {Zhang}}]{liu:23}
{Liu}, Q., {Abraham}, R., {Martin}, P.~G., {et~al.} 2023, \apj, 953, 7,
  \dodoi{10.3847/1538-4357/acdee3}

\bibitem[{{Lokhorst} {et~al.}(2022{\natexlab{a}}){Lokhorst}, {Abraham},
  {Pasha}, {van Dokkum}, {Chen}, {Miller}, {Danieli}, {Greco}, {Zhang},
  {Merritt}, \& {Conroy}}]{lokhorst:22}
{Lokhorst}, D., {Abraham}, R., {Pasha}, I., {et~al.} 2022{\natexlab{a}}, \apj,
  927, 136, \dodoi{10.3847/1538-4357/ac50b6}

\bibitem[{{Lokhorst} {et~al.}(2022{\natexlab{b}}){Lokhorst}, {Abraham},
  {Pasha}, {van Dokkum}, {Chen}, {Miller}, {Danieli}, {Greco}, {Zhang},
  {Merritt}, \& {Conroy}}]{Lokhorst:2022}
---. 2022{\natexlab{b}}, \apj, 927, 136, \dodoi{10.3847/1538-4357/ac50b6}

\bibitem[{{Lomel\'i-Huerta} {et~al.}(2022){Lomel\'i-Huerta}, {Rivera-Caicedo},
  {De-la-Torre}, {Acevedo-Ju\'arez}, {Cepeda-Morales}, \&
  {Avila-George}}]{lomeli:22}
{Lomel\'i-Huerta}, J.~R., {Rivera-Caicedo}, J.~P., {De-la-Torre}, M., {et~al.}
  2022, PeerJ Comput Sci., 8, e979, \dodoi{10.7717/peerj-cs.979}

\bibitem[{{Marois} {et~al.}(2006){Marois}, {Lafreni{\`e}re}, {Doyon},
  {Macintosh}, \& {Nadeau}}]{marois:06}
{Marois}, C., {Lafreni{\`e}re}, D., {Doyon}, R., {Macintosh}, B., \& {Nadeau},
  D. 2006, \apj, 641, 556, \dodoi{10.1086/500401}

\bibitem[{{Montes} \& {Trujillo}(2018)}]{montes:18}
{Montes}, M., \& {Trujillo}, I. 2018, \mnras, 474, 917,
  \dodoi{10.1093/mnras/stx2847}

\bibitem[{{Neill} {et~al.}(2023){Neill}, {Matuszewski}, {Martin}, {Brodheim},
  \& {Rizzi}}]{neill:23}
{Neill}, D., {Matuszewski}, M., {Martin}, C., {Brodheim}, M., \& {Rizzi}, L.
  2023, {KCWI\_DRP: Keck Cosmic Web Imager Data Reduction Pipeline in Python},
  Astrophysics Source Code Library, record ascl:2301.019.
\newblock \doeprint{2301.019}

\bibitem[{{Newman} \& {Sproull}(1979)}]{newman:79}
{Newman}, W.~M., \& {Sproull}, R.~F. 1979, {Principles of Interactive Computer
  Graphics (2nd edition)} (New York: McGraw-Hill)

\bibitem[{{Pasha} \& {Miller}(2023)}]{Pasha:2023}
{Pasha}, I., \& {Miller}, T.~B. 2023, The Journal of Open Source Software, 8,
  5703, \dodoi{10.21105/joss.05703}

\bibitem[{{Pasha} {et~al.}(2021){Pasha}, {Lokhorst}, {van Dokkum}, {Chen},
  {Abraham}, {Greco}, {Danieli}, {Miller}, {Lippitt}, {Polzin}, {Shen}, {Keim},
  {Liu}, {Merritt}, \& {Zhang}}]{Pasha:2021}
{Pasha}, I., {Lokhorst}, D., {van Dokkum}, P.~G., {et~al.} 2021, \apjl, 923,
  L21, \dodoi{10.3847/2041-8213/ac3ca6}

\bibitem[{{Peng} {et~al.}(2002){Peng}, {Ho}, {Impey}, \& {Rix}}]{galfit}
{Peng}, C.~Y., {Ho}, L.~C., {Impey}, C.~D., \& {Rix}, H.-W. 2002, \aj, 124,
  266, \dodoi{10.1086/340952}

\bibitem[{{Popowicz} {et~al.}(2013){Popowicz}, {Kurek}, \&
  {Filus}}]{popowicz:13}
{Popowicz}, A., {Kurek}, A.~R., \& {Filus}, Z. 2013, \pasp, 125, 1119,
  \dodoi{10.1086/673179}

\bibitem[{{Popowicz} \& {Smolka}(2015)}]{popowicz:15}
{Popowicz}, A., \& {Smolka}, B. 2015, \mnras, 452, 809,
  \dodoi{10.1093/mnras/stv1320}

\bibitem[{{Price-Whelan} \& {Astropy Collaboration}(2018)}]{astropy2018}
{Price-Whelan}, A.~M., \& {Astropy Collaboration}. 2018, \aj, 156, 123,
  \dodoi{10.3847/1538-3881/aabc4f}

\bibitem[{{Sakurai} \& {Shin}(2001)}]{sakurai:01}
{Sakurai}, T., \& {Shin}, J. 2001, \pasj, 53, 361,
  \dodoi{10.1093/pasj/53.2.361}

\bibitem[{{Saydjari} \& {Finkbeiner}(2022)}]{saydjari:22}
{Saydjari}, A.~K., \& {Finkbeiner}, D.~P. 2022, \apj, 933, 155,
  \dodoi{10.3847/1538-4357/ac6875}

\bibitem[{{Tody}(1986)}]{iraf1}
{Tody}, D. 1986, in \procspie, Vol. 627, Instrumentation in astronomy VI, ed.
  D.~L. {Crawford}, 733, \dodoi{10.1117/12.968154}

\bibitem[{{Tody}(1993)}]{iraf2}
{Tody}, D. 1993, in Astronomical Society of the Pacific Conference Series,
  Vol.~52, Astronomical Data Analysis Software and Systems II, ed. R.~J.
  {Hanisch}, R.~J.~V. {Brissenden}, \& J.~{Barnes}, 173

\bibitem[{{Ulyanov} {et~al.}(2020){Ulyanov}, {Vedaldi}, \&
  {Lempitsky}}]{ulyanov:20}
{Ulyanov}, D., {Vedaldi}, A., \& {Lempitsky}, V. 2020, Int.\ J.\ Comput.\ Vis.,
  128, 1867, \dodoi{10.1007/s11263-020-01303-4}

\bibitem[{{van Dokkum} {et~al.}(2020){van Dokkum}, {Lokhorst}, {Danieli}, {Li},
  {Merritt}, {Abraham}, {Gilhuly}, {Greco}, \& {Liu}}]{dokkum:mrf}
{van Dokkum}, P., {Lokhorst}, D., {Danieli}, S., {et~al.} 2020, \pasp, 132,
  074503, \dodoi{10.1088/1538-3873/ab9416}

\bibitem[{{van Dokkum}(2001)}]{dokkumc:01}
{van Dokkum}, P.~G. 2001, \pasp, 113, 1420.
\newblock
  \url{http://adsabs.harvard.edu/cgi-bin/nph-bib_query?bibcode=2001PASP..113.1420V&db_key=AST}

\bibitem[{Walt {et~al.}(2011)Walt, Colbert, \& Varoquaux}]{numpy}
Walt, S. v.~d., Colbert, S.~C., \& Varoquaux, G. 2011, Computing in Science and
  Engg., 13, 22, \dodoi{10.1109/MCSE.2011.37}

\bibitem[{{Zhang} \& {Bloom}(2020)}]{zhang:20}
{Zhang}, K., \& {Bloom}, J.~S. 2020, \apj, 889, 24,
  \dodoi{10.3847/1538-4357/ab3fa6}

\end{thebibliography}

\end{document}